\documentclass[12pt,letterpaper]{iopart}

\usepackage{iopams}
\usepackage{hyperref}
\usepackage{graphicx}
\usepackage{color}
\usepackage{bm}

\begin{document}
	
\title[Origin of level attraction in a microwave cavity]{Microscopic origin of level attraction for a coupled magnon-photon system in a microwave cavity}

\author{Igor Proskurin$^{1,2}$, Rair Mac\^{e}do$^{3}$, Robert L Stamps$^{1}$}
\address{$^{1}$ Department of Physics and Astronomy, University of Manitoba, Winnipeg, Manitoba R3T 2N2, Canada}
\address{$^{2}$ Institute of Natural Sciences and Mathematics, Ural Federal University, Ekaterinburg 620002, Russia}
\address{$^{3}$ James Watt School of Engineering, Electronics \& Nanoscale Engineering Division, University of Glasgow, Glasgow G12 8QQ, UK}
\ead{Igor.Proskurin@umanitoba.ca}

\begin{abstract}
We discuss various microscopic mechanisms for level attraction in a hybridized magnon-photon system of a ferromagnet in a microwave cavity.  The discussion is based upon the electromagnetic theory of continuous media where the effects of the internal magnetization dynamics of the ferromagnet are described using dynamical response functions. This approach is in agreement with quantized multi-oscillator models of coupled photon-magnon dynamics.   We demonstrate that to provide the attractive interaction between the modes, the effective response functions should be  diamagnetic.  Magneto-optical coupling is found to be one mechanism for the effective diamagnetic response, which is proportional to photon number.  A dual mechanism based on the Aharonov-Casher effect is also highlighted, which is instead dependent on magnon number.
\end{abstract}

\noindent{\it Keywords\/}: level attraction, microwave cavities, magneto-optical interactions, cavity optomagnonics

\section{Introduction: mode attraction in microwave cavities}

Microwave cavity resonators are useful for coupling microwave photons to various excitations including mechanical \cite{Aspelmeyer2014}, acoustic \cite{Eichenfield2009}, and magnetic degrees of freedom \cite{Harder2018}.  In magnetism, hybridization between cavity modes and ferromagnetic resonances in the form of cavity magnon-polaritons has been demonstrated and studied in a number of systems \cite{Tabuchi2014,Zhang2014,Bai2014,Cao2015,Yao2015,Zhang2015,Harder2016,Maier2016,Yao2018}.  Cavity optomagnonics was proposed recently with analogues to optomechanics \cite{Zhang2016a,Osada2016,Kusminskiy2016,Liu2016}.  In optomagnonics, cavity photons couple to magnetic excitations via the magneto-optical interactions, which allows  characterization using a well-established optomechanical Hamiltonian \cite{Kusminskiy2016,Liu2016}, and, as a consequence, one should be able to observe optomechanical effects in magnetic systems \cite{Sharma2018}.  Various phenomena have been proposed for coupled magnon-photon systems inside microwave cavities including Brillouin-scattering-induced transparency in whispering gallery resonators  \cite{Dong2015,Zhang2016,Sharma2017}, collective dynamics of spin textures \cite{Graf2018,Proskurin2018}, and photon-mediated nonlocal interactions \cite{Lambert2016,Zare2018,Bai2018,Johansen2018}.

Recently, for certain positions of a magnetic sample a mode attraction regime has been observed within a cylindrical \cite{Harder2018a} or a planar cavity \cite{Yang2019}.  Mode attraction is a general feature know already to be possible in dynamic optomechanical systems \cite{Bernier2018}.  Naturally, it requires a certain instability mechanism being provided, which in optomechanical context is realized via the negative frequency in the effective Hamiltonian of a driven system \cite{Bernier2018}. In contrast to usual mode hybridization, mode attraction is characterized by a region where the real parts of the eigenfrequencies coalesce marked by the exceptional points where the eigenmodes collapse \cite{Seyranian2005,Heiss2012}.  In Ref.~\cite{Harder2018a}, the level attraction for cavity magnon-polaritons was interpreted as a manifestation of Lenz's law in a phenomenological electrical circuit model.  The mechanism for level attraction between strongly interacting spin-photon excitations was also proposed for a system with two driving terms with a phase offset \cite{Grigoryan2018}, which was recently realized in a re-entrant cavity resonator using two separate drives \cite{Boventer2019, Boventer2019a}.

The purpose of this paper is to discuss possible microscopic material mechanisms for mode attraction in ferromagnetic systems. Most of these mechanisms can be thought of in terms of an effective diamagnetic response of the system.  To show this, we will use the approach based on the Maxwell's equations in dispersive media, where the internal magnetization dynamics is taken into account through response functions and constitutive relations \cite{Zare2015}.

\begin{figure}
	\centerline{\includegraphics[width=0.875\textwidth]{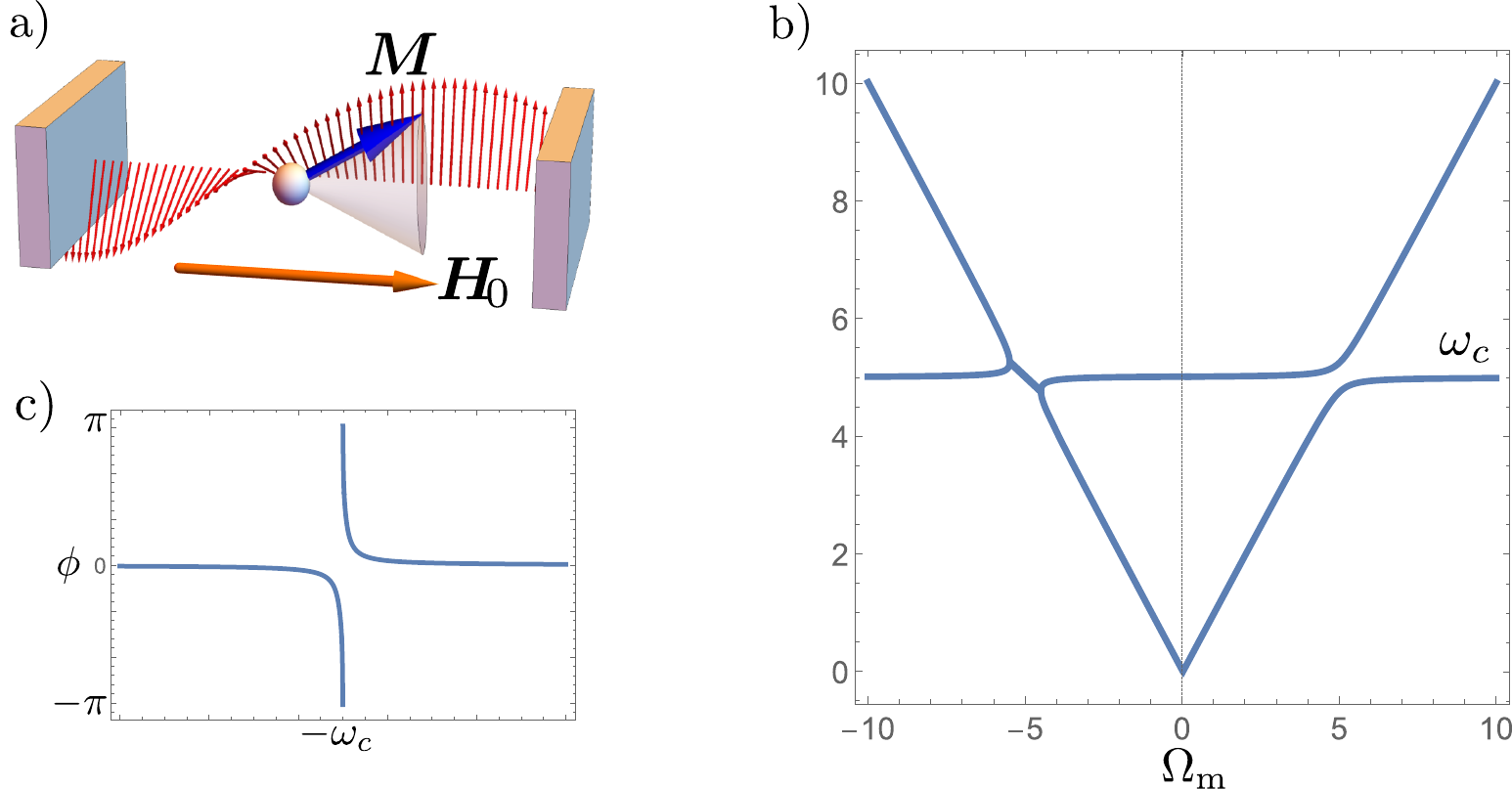}}
	\caption{a) Schematic picture of a small ferromagnetic sample interacting with circularly polarized electromagnetic wave inside a microwave cavity. b) Energy dispersion for the Hamiltonian (\ref{Htot}) as a function of $\Omega_{\mathrm{m}}$ showing level repulsion near $\omega_{c}$ and level attraction at $-\omega_{c}$. c) Phase shift in the equation (\ref{phi}) as a function of $\omega$ demonstrating $2\pi$ jump near $-\omega_{c}$.}
	\label{fig1}
\end{figure}

Before going to specific details, we illustrate the essential ideas with a three coupled oscillator model.  Such a model is obtained if we take a small
magnetic specimen with the magnetization $\bm{M}(t)$ in a static magnetic field $\bm{H}_{0}$ interacting with the microwave field $\bm{h}(t)$ of the cavity, as schematically shown in Fig.~\ref{fig1} (a). This model shows the level attraction in the unstable regime where the magnetization is opposite to the magnetic field, which formally corresponds to a negative magnon frequency. Such configuration is, of course, unphysical.  However, it illustrates a general rule that the mode attraction is related to an instability in the system.  In the linear approximation for magnetization dynamics, the total interacting Hamiltonian can be written as (see \ref{AppA} for details)
\begin{equation} \label{Htot}
\mathcal{H} = \hbar\omega_{c} \left(a_{L}^{\dag}a_{L} 
+ a_{R}^{\dag}a_{R}\right) + \hbar\Omega_{\mathrm{m}} b^{\dag} b 
+ \hbar g_{0} \left[(a_{R} + a_{L}^{\dag})b + \mbox{H.c.}\right] + \mathcal{H}_{\mathrm{drv}},
\end{equation}
where $a^{\dag}$ ($a$) is the creation (annihilation) operator for two degenerated cavity photon modes with right ($R$) and left ($L$) polarizations and the frequency $\omega_{c}$, $b$ and $b^{\dag}$ describe the magnon oscillator with the frequency $\Omega_{\mathrm{m}}$, $g_{0}$ denotes the magnon-photon coupling paramenter, and the last term is the external driving energy at the frequency $\omega$, $\mathcal{H}_{\mathrm{drv}} = i\hbar\sqrt{\kappa_{\mathrm{ex}}}
[\alpha_{\mathrm{in}}a_{R}\exp(-i\omega t) 
- \alpha_{\mathrm{in}}^{*}a_{R}^{\dag}\exp(i\omega t)]$ with the amplitude  $\alpha_{\mathrm{in}}$ and coupling parameter $\kappa_{\mathrm{ex}}$.

The qualitative behavior of this model can be understood directly from the Hamiltonian (\ref{Htot}).  For $\Omega_{\mathrm{m}} \approx \omega_{c}$, the dominant contribution comes from the interaction term proportional to $a_{L}^{\dag}b + a_{L}b^{\dag}$  that describes a hybridization between the left polarized electromagnetic wave and the magnon mode precessing in the same direction, while $a_{R}$ remains decoupled \cite{Akhiezer1968}.  More interesting behavior takes place for $\Omega_{\mathrm{m}} \approx -\omega_{c}$, where the term $a_{R}b + a_{R}^{\dag}b^{\dag}$ is dominant, which is know as the `two-mode squeezing' regime in optomechanics \cite{Aspelmeyer2014}.  In this case, the level attraction regime is observed between the magnon and the right-polarized cavity mode in the region $(\Omega_{\mathrm{m}} + \omega_{c})^{2} < 4g^{2}$ \cite{Bernier2018}. The whole picture is summarized in Fig.~\ref{fig1} (b).

Another salient feature in the mode attraction regime of the three oscillator model appears with varying the frequency of the drive around the point $\omega = -\omega_{c}$ when all three modes are degenerated, $\omega_{c}=\Omega_{\mathrm{m}}$.  This appears as a phase shift between the driving field $\alpha_{\mathrm{in}}$ and the driven cavity mode $a_{R}$.  If we introduce the dissipation in the magnon channel, $\Omega_{\mathrm{m}} - i\kappa_{\mathrm{m}}$, and neglect cavity dissipation, phase shift between the driving term and the response field is estimated as
\begin{equation} \label{phi}
\tan \phi = \frac{\kappa_{\mathrm{m}}(\omega^{2} - \omega_{c}^{2})}{(\omega^{2} - \omega_{c}^{2})(\omega + \Omega_{\mathrm{m}}) + 2g_{0}^{2}\omega_{c}},
\end{equation} 
which contains contributions from two degenerated cavity modes with different polarizations and the magnon mode.  When all modes are degenerated,  $\Omega_{\mathrm{m}} = -\omega_{c}$, these two contributions provide the phase shifts of $+\pi$ and $-\pi$ around $\omega = -\omega_{c}$ that results in a characteristic $2\pi$ phase jump as shown in Fig.~\ref{fig1} (c).  We believe that similar mechanism may be behind the phase shift reported in Ref.~\cite{Harder2018a}. We note, however, that within this illustration, $2\pi$ phase shift occurs at negative frequencies, whereas in the physical region $\omega \approx \omega_{c}$ it equals to $\pi$. Below, we will consider the response functions that can bring $2\pi$ phase jump to positive $\omega$.

As already mentioned, the condition $\Omega_{\mathrm{m}} \approx -\omega_{c}$ represents an unstable configuration where $\bm{M}$ and $\bm{H}_{0}$ are antiparallel. Another possible scenario would require fully dissipative imaginary coupling constant $g_{0}$. Either possibility explains why level attraction was not observed previously for cavity magnon-polaritons. In this paper, we discuss physical mechanisms that are not fully dissipative, but instead arise from effective `diamagnetic' responses where the reaction of the system is towards compensating the applied excitation.  This can be somehow from the cavity itself, as proposed in Refs.~\cite{Harder2018a,Yang2019}, or through some mechanism in the sample response.  We consider here possible mechanisms first via the electromagnetic pressure on the sample through electro-optics, and second via the magnon pressure on the cavity through the Aharonov-Bohm mechanism.

In the following we construct a classical electromagnetic theory for the measured responses in terms of cavity and sample impedances. This allows us to define a tensor magnetic response for cavity-magnon coupling through either magnetic or electro-optical mechanisms. We demonstrate that magneto-optical coupling, or more specifically, the inverse Faraday effect, provides one possibility for the effective diamagnetic response, which being quantized, leads to a linearized optomagnonic Hamiltonian \cite{Kusminskiy2016,Liu2016}. Another possible scenario for the attractive regime comes from the Aharonov-Casher effect \cite{Aharonov1984}.  As we discuss at the end, this effect can be considered as dual to the magneto-optical coupling. In both scenarios, the magnon-photon interaction is driven by the cavity electric field.

\section{Electromagnetic theory and diamagnetic response}

A theory sufficiently complex to allow for different placements within a cavity in an actual experiment can be obtained using classical electrodynamics. This is most naturally formed in terms of impedances, and provides a direct connection to the scattering matrix element $S_{21}(\omega)$ typically measured in experiment. Frequency shifts and phase can be obtained from the complex scattering elements. These are obtained using standard classical electromagnetic theory in terms of energies $W_e$ and $W_m$, defined over the sample volume $V_S$ as
\numparts
\begin{eqnarray}
\label{we}
W_e &= i \omega \int d^3r [\hat{\varepsilon}(\omega)-\hat{\varepsilon}_0] \bm{E}\cdot \bm{E}_0^*, \\
\label{wb}
W_m &= i \omega \int d^3r [\hat{\mu}(\omega)-\hat{\mu}_0] \bm{H}\cdot \bm{H}_0^*,
\end{eqnarray}
\endnumparts
where the complex electric permittivity, $\hat{\varepsilon} = 1 + \hat{\alpha}$, and magnetic permeability, $\hat{\mu} = 1 + \hat{\chi}$ are different from those for the free space due to the presence of the sample. The cavity electric and magnetic fields satisfy Maxwell's equations in free space and are $\bm{E}_0$ and $\bm{H}_0$. The fields perturbed by the sample are $\bm{E}$ and $\bm{H}$.  Together the energies provide the sample contribution to the measured impedance $Z_S=W_e - W_m$ \cite[\S6.10]{Jackson1998}.  The perturbative impedance $Z_{S}$ can then be simply added to a cavity impedance $Z_{c}$ with parameters determined by unloaded measurements.

The required properties needed to construct $Z_{S}$ from the response functions for the loaded cavity can be understood in terms of the electromagnetic theory of continuous media.  In this approach, we ignore at the beginning the finite sample size and describe the field inside the cavity by the Maxwell's equations 
\numparts
\begin{eqnarray}
\bm{\nabla} \times \bm{E} &= -\partial_{t} \left(\mu_{0}\hat{\mu}\bm{H}\right), \quad \bm{\nabla} \cdot \left(\varepsilon_{0}\hat{\varepsilon}\bm{E}\right) = 0, \\
\bm{\nabla} \times \bm{H} &= \partial_{t} \left(\varepsilon_{0}\hat{\varepsilon}\bm{E}\right), \quad \bm{\nabla} \cdot \left(\mu_{0}\hat{\mu}\bm{H}\right) = 0,
\end{eqnarray}
\endnumparts
By expanding the fields, $\bm{E} = \bm{E}_{0} + \delta\bm{E}$ and $\bm{H} = \bm{H}_{0} + \delta\bm{H}$, where $\delta\bm{E}$ and $\delta\bm{H}$ are related to the sample, we rewrite the equations above in the form
\numparts
\begin{eqnarray}\label{Mx1}
\bm{\nabla} \times \delta \bm{E} + \partial_{t}\left(\mu_{0}\hat{\mu}\delta\bm{H}\right) & = -\partial_{t} \left(\mu_{0}\hat{\chi}\bm{H}_{0}\right), \\\label{Mx2}
\bm{\nabla} \times \delta \bm{H} - \partial_{t}\left(\varepsilon_{0}\hat{\varepsilon}\delta\bm{E}\right) & = \partial_{t} \left(\varepsilon_{0}\hat{\alpha}\bm{E}_{0}\right),
\end{eqnarray}
\endnumparts
where the free space cavity field can be considered as driving terms, which provide the response determined by the material functions of the medium.

Let us consider a situation where the plane electromagnetic wave is propagating in a homogeneous dispersive birefringent medium along the gyrotropic axis taken as $z$ direction.  In this case, we take magnetic susceptibility tensor in the form
\begin{equation}
\hat{\chi}(\omega) = 
\left(
\begin{array}{ccc}
	\chi_{1}(\omega)   & i\chi_{2}(\omega) & 0 \\
	-i\chi_{2}(\omega) & \chi_{1}(\omega)  & 0 \\
	0                  & 0                 & 0
\end{array}
\right),
\end{equation}
and $ \hat{\varepsilon} = 1 $ for simplicity. Transforming (\ref{Mx1}) and (\ref{Mx2}) to the $\omega$-domain, and replacing $\partial_{z}$ by $i\omega_{c}/c$, where $\omega_{c}$ is some characteristic frequency of the cavity,  we obtain
\begin{equation}\fl
\label{eom1}
\left(
\begin{array}{cccc}
	\mu_{0}\omega (1 + \chi_{1}) & i\mu_{0}\omega\chi_{2}       & 0                     & \omega_{c}/c           \\
	-i\mu_{0}\omega\chi_{2}      & \mu_{0}\omega (1 + \chi_{1}) & -\omega_{c}/c         & 0                      \\
	0                            & -\omega_{c}/c                & \varepsilon_{0}\omega & 0                      \\
	\omega_{c}/c                 & 0                            & 0                     & \varepsilon_{0} \omega
\end{array}
\right)
\left(
\begin{array}{c}
\delta H_{x} \\ \delta H_{y} \\ \delta E_{x} \\ \delta E_{y}
\end{array}
\right)
=
-\mu_{0}\omega
\left(
\begin{array}{cc}
\hat{\chi}(\omega)\bm{H}_{0} \\
0
\end{array}
\right).
\end{equation}
If we include dissipation into account, $\chi_{1}$ and $\chi_{2}$ become complex. In this situation, the equation above has the exceptional points at the frequencies, which satisfy the relation $\chi_{2}(\omega) - \chi_{1}(\omega) = 2i\omega_{c}/\omega$, where the matrix on the left hand side does not have a diagonal form \cite{Heiss2012}.

Calculating the circular components of the response magnetic field from equation (\ref{eom1}), we obtain
\begin{equation}
\delta H^{(\pm)}(\omega) = -\frac{\left[\chi_{1}(\omega) \pm \chi_{2}(\omega)\right]\omega^{2}\delta H_{0}^{(\pm)}}{\left[1 + \chi_{1}(\omega) \pm \chi_{2}(\omega)\right]\omega^{2} - \omega_{c}^{2}},
\end{equation}
where $\delta H^{(\pm)} = \delta H_{x} \pm i\delta H_{y}$, which is directly related to the impedance of the loaded cavity via equations (\ref{we}, \ref{wb}).  If we take the components of the susceptibility for the ferromagnetic resonance \cite{Akhiezer1968}, $\chi_{1}(\omega) = \gamma M_{s}\Omega_{\mathrm{m}}/(\Omega_{\mathrm{m}}^{2}-\omega^{2})$ and $\chi_{2}(\omega) = \gamma M_{s}\omega/(\Omega_{\mathrm{m}}^{2}-\omega^{2})$, we obtain a result consistent with equations (\ref{res})--(\ref{res1})
\begin{equation}
\delta H^{(\pm)}(\omega) =-\frac{\gamma M_{s}\omega^{2}\delta H_{0}^{(\pm)}}{(\omega^{2} - \omega_{c}^{2})(\Omega_{\mathrm{m}} \mp \omega) + \gamma M_{s} \omega^{2}},
\end{equation}
which shows level repulsion behavior around $\omega \approx \Omega_{\mathrm{m}}$. To obtain level attraction instead of repulsion, we formally need a `diamagnetic' response here, which corresponds to $\gamma M_{s} < 0$.

\subsection{Inverse Faraday effect mechanism}

From the physical point of view, the effective resonant diamagnetic response may be realized if we consider nonlinear effects in light-matter interactions.  One particular example is the inverse Faraday effect \cite[\S101]{Landau2013}, where circular polarized components of the electric field create an effective magnetic field $\bm{H}^{\mathrm{eff}}(t) = i \varepsilon_{0} f \bm{\mathcal{E}}(t) \times \bm{\mathcal{E}}^{*}(t)/4$, where $\bm{\mathcal{E}}(t)$ is the electric field amplitude in the complex representation, $\bm{E} = (\bm{\mathcal{E}}  + \bm{\mathcal{E}}^{*})/2$, and $f$ is the material dependent parameter.  The effective magnetic field is able to drive the dynamics of the magnetization towards the resonance, which can be taken into account in  Maxwell's equations through the modulation of the electric permittivity.  Compared to the usual magneto-dipole coupling between the magnetization and the cavity magnetic field, we expect that this mechanism is dominant in the nodes of the cavity modes, where the magnetic field is small, while the electric field is maximal.  By moving a small specimen inside the cavity, we can tune the relative strength of different coupling mechanisms, which changes the hybridization picture \cite{Harder2018a}.

For illustration, we consider the experimental setup shown in Fig.~\ref{fig2} (a), where the cavity electromagnetic wave is excited along the $x$-direction perpendicular to the saturation axis of the magnetization, which is taken as $z$-axis.  In this case, the displacement field generated by the inverse Faraday effect is given by $\delta D_{i}(t) = i \varepsilon_{0} f \epsilon_{ijx} m_{x}(t) E_{j}(t)$, where $\epsilon_{ijk}$ denotes the Levi-Civita tensor.  If we consider only linear terms in the fluctuating parts of the fields, and transform to the Fourier space, the effective response that takes into account resonant magnetization behavior can be estimated as 
\begin{equation} \label{dia}
\fl
\left(
\begin{array}{c}
\delta \mathcal{D}_{y}(\omega) \\
\delta \mathcal{D}_{z}(\omega)
\end{array}
\right)
= - \frac{\varepsilon_{0}^{2}f^{2}}{4} \chi_{xx}(\omega-\omega_{\mathrm{L}})
\left(
\begin{array}{cc}
|\mathcal{E}_{0z}|^{2} & -\mathcal{E}_{0y}^{*} \mathcal{E}_{0z}  \\
-\mathcal{E}_{0z}^{*} \mathcal{E}_{0y} & |\mathcal{E}_{0y}|^{2}
\end{array}
\right)
\left(
\begin{array}{c}
\delta \mathcal{E}_{y}(\omega) \\
\delta \mathcal{E}_{z}(\omega)
\end{array}
\right),
\end{equation}
where $\bm{D}(\omega) = [\bm{\mathcal{D}}(\omega) + \bm{\mathcal{D}}^{*}(\omega)]/2$, $ \bm{\mathcal{E}}_{0} $ and $\omega_{\mathrm{L}}$ are the amplitude and the frequency of the driving field, and we neglected the terms with $\bm{\mathcal{E}}(\omega \pm 2 \omega_{\mathrm{L}})$. The details can be found in \ref{AppB}.  Equation (\ref{dia}) shows that in the linear approximation the inverse Faraday effect provides an effective dielectric resonant permittivity proportional to the intensity of the driving field and consistent with the macroscopic Lenz's effect. However, we note that such power dependence of the parameters has not been observed in recent experiments \cite{Harder2018a, Yang2019}, which means that these systems require a different microscopic explanation.  We expect that the results of this section will be relevant to optomagnonic systems discussed in Refs.~\cite{Kusminskiy2016,Liu2016}.

If we apply the formalism of equations (\ref{Mx1}, \ref{Mx2}) together with the effective permittivity in (\ref{dia}), excluding the magnetic field, we obtain the following equations for the fluctuating cavity electric field
\begin{equation}\fl\nonumber
\left(
\begin{array}{cc}
\omega_{c}^{2} - \omega^{2} + |g_{z}|^{2} \chi_{xx}(\bar{\Delta})\omega^{2} & -g_{y}^{*}g_{z} \chi_{xx}(\bar{\Delta}) \omega^{2} \\
-g_{y}g_{z}^{*} \chi_{xx}(\bar{\Delta}) \omega^{2} & \omega_{c}^{2} - \omega^{2} - \chi_{yy}(\omega)\omega^{2} + |g_{y}|^{2} \chi_{xx}(\bar{\Delta})\omega^{2}
\end{array}
\right)
\left(
\begin{array}{c}
\delta \mathcal{E}_{y} \\ \delta \mathcal{E}_{z}
\end{array}
\right)
= 0,
\label{eom2}
\end{equation}
where we have introduced effective coupling constants $g_{i} = \varepsilon_{0}^{1/2} f \mathcal{E}_{0i}/4$ ($i=y,z$), and where $\bar{\Delta} = \omega - \omega_{\mathrm{L}}$ with $\omega$ being the frequency of the probe field.  The equation above shows that the magneto-optical coupling to a dynamic magnetization, in the linear approximation,  is equivalent to an effective `diamagnetic' response.  This becomes apparent if we consider the equation for $\delta\mathcal{E}^{z}$ where the magneto-optical contribution competes with the ordinary ferromagnetic resonance [$\sim \chi_{yy}(\omega)$] from the $y$-component of the fluctuating magnetic field.

\subsection{Magneto-optical coupling mechanism}

\begin{figure}
	\centerline{\includegraphics[width=0.65\textwidth]{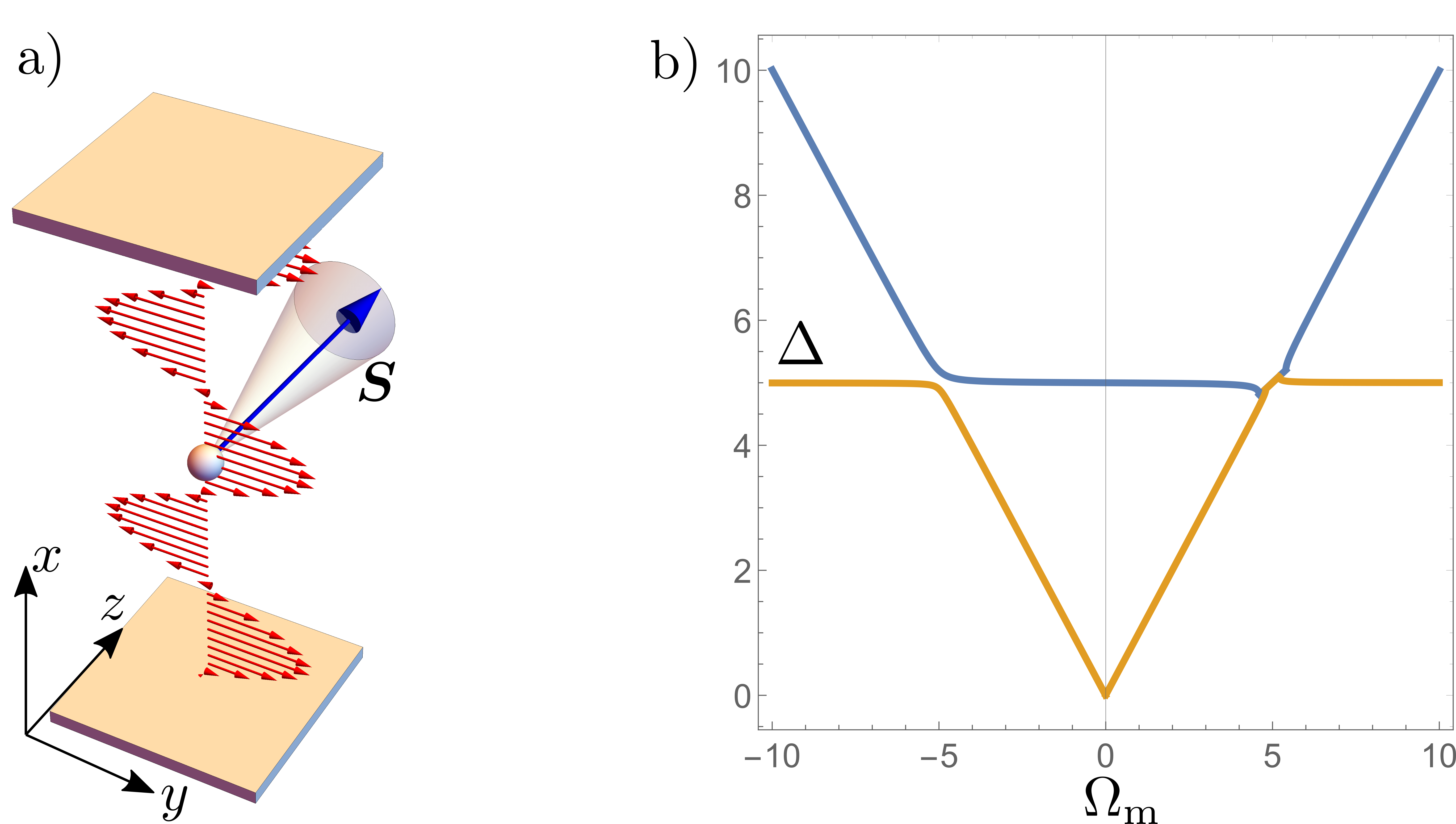}}
	\caption{a) Experimental geometry for the magneto-optical mechanism of the mode attraction. b) Energy dispersion (\ref{omp}, \ref{omm}) as a function of $\Omega_{\mathrm{m}}$.}
	\label{fig2}
\end{figure}

A possible origin for the effective diamagnetic contribution in (\ref{dia}) and (\ref{eom2}) can be understood from a quantized optomagnonic Hamiltonian \cite{Kusminskiy2016, Liu2016}.  For this purpose, we consider a spin $\bm{S}=\bm{S}\delta(\bm{r}-\bm{r}_{0})$ coupled to the electric field through the magneto-optical coupling
\begin{equation}
\mathcal{H}_{\mathrm{mo}} = -\frac{\varepsilon_{0}}{4} \int \delta \varepsilon_{ij} \mathcal{E}_{i} \mathcal{E}_{j}^{*} d^{3}r,
\end{equation}
where $ \delta \varepsilon_{ij} = if\epsilon_{ijk}S_{k} $ describes the inverse Faraday effect with $f=2c\theta_{F}/(\sqrt{\varepsilon}\omega S)$ \cite{Kusminskiy2016}, where $\theta_{F}$ is the Faraday rotation angle, $ c $ is the velocity of light, $\omega$ is the frequency of the electromagnetic wave, and $\varepsilon$ is the electric permittivity of the medium. In what follows, we take the electric field in the form of cavity standing waves along the $x$ axis, $\bm{\mathcal{E}}(x)=\sum_{\lambda=L,R}[\hbar\omega_{c}/(\varepsilon_{0}V)]^{1/2}\sin(\pi n x_{0}/L_{x}) \bm{e}_{\lambda}a_{\lambda}$, where $a_{\lambda}$ is the cavity photon annihilation operator, $\bm{e}_{\lambda} = (0,\lambda,i)/\sqrt{2}$ is the polarization vector for the circularly polarized wave with the left ($L=-1$) or right ($R=1$) polarization $\lambda$, $n$ is the cavity mode index, $V$ is the volume of the cavity, $L_{x}$ is the size of the cavity along the $ x $ direction, and $x_{0}$ denotes the position of the sample. We take the $ z $-axis as a quantization axis for the spin $ \bm{S} $, as shown in Fig.~\ref{fig2} (a), so that the magneto-optical coupling becomes
\begin{equation}
\label{Hmo}
\mathcal{H}_{\mathrm{mo}} =  
\hbar g_{0} (a_{L}^{\dag}a_{L} - a_{R}^{\dag}a_{R})(b + b^{\dag}),
\end{equation}
where $g_{0} = c\theta_{F} \xi/\sqrt{2\varepsilon S}$ is the magnon-photon coupling parameter at the node where the electric field is at the maximum ($\xi \lesssim 1$ is a geometric factor). For $\mu$m sized yttrium iron garnet sample at the optical wavelength $\sim 1$~$\mu$m, we can estimate $\theta_{F} = 200^{\circ}$~cm$^{-1}$ \cite[App.~A]{Stancil2012}, $ \varepsilon \approx 5 $ and $S=10^{10}$, which gives $g_{0} \approx 10^{5}$~Hz \cite{Kusminskiy2016}.

The interaction energy in (\ref{Hmo}) is the same as in optomechanical applications.  In the rotating wave approximation (RWA), the total Hamiltonian can be written as
\begin{equation}\fl
\label{Htot1}
\mathcal{H} = -\hbar\Delta (a_{1}^{\dag}a_{1} + a_{2}^{\dag}a_{2}) + \hbar \Omega_{\mathrm{m}} b^{\dag} b + i\hbar g_{0} (a_{2}^{\dag}a_{1} - a_{1}^{\dag}a_{2})(b + b^{\dag}) + i\hbar\sqrt{\kappa_{\mathrm{ex}}}(\alpha a_{1}^{\dag} - i\alpha^{*} a_{1}),
\end{equation}
where $ \alpha$ is the amplitude of the driving field, and we have transformed to the linearly polarized basis, $a_{L} = (a_{1} - i a_{2})/\sqrt{2}$ and $a_{R} = (a_{1} + i a_{2})/\sqrt{2}$.  The last term in this expression is the external driving, which becomes time independent under the RWA \cite{Aspelmeyer2014}.  Compared to the magneto-dipole interaction in (\ref{Htot}), the optomechanical coupling in RWA leads to the detuning parameter $\Delta = \omega_{\mathrm{L}} - \omega_{c}$ that may have an arbitrary sign depending on the ratio of the driving frequency, $\omega_{\mathrm{L}}$, to $\omega_{c}$.

The interaction part of the Hamiltonian (\ref{Htot1}) can be linearized as follows:
\begin{equation}
a_{2}^{\dag}a_{1} - a_{1}^{\dag}a_{2} \approx \sqrt{n_{2}} (a_{1} - a_{1}^{\dag}) - \sqrt{n_{1}} (a_{2} - a_{2}^{\dag}),
\end{equation}
where $n_{1}$ and $n_{2}$ denote the average numbers of cavity photons.  The linearized Hamiltonian is written as
\begin{eqnarray}
\nonumber
\mathcal{H}_{\mathrm{lin}} &=& -\hbar\Delta (a_{1}^{\dag}a_{1} + a_{2}^{\dag}a_{2}) + \hbar \Omega_{\mathrm{m}} b^{\dag} b + i\hbar g_{1} (a_{1} - a_{1}^{\dag})(b + b^{\dag})
\\  \label{Hlin}
&& - i\hbar g_{2} (a_{2} - a_{2}^{\dag})(b + b^{\dag}) + i\hbar\sqrt{\kappa_{\mathrm{ex}}}(\alpha a_{1}^{\dag} - i\alpha^{*} a_{1}),
\end{eqnarray}
where we have introduced renormalized coupling parameters $ g_{1} = g_{0}\sqrt{n_{2}} $ and $ g_{2} = g_{0}\sqrt{n_{1}} $. In a situation where the driving field is coupled only to the cavity photon of one polarization, we would normally expect $g_{2} \gg g_{1}$.

The Heisenberg equations of motion for the Hamiltonian (\ref{Hlin}) can be written in the matrix form:
\begin{equation}\fl
\label{eom3}
-i\partial_{t}
\left(
\begin{array}{c}
a_{1} \\ a_{1}^{\dag} \\ a_{2} \\ a_{2}^{\dag} \\ b \\ b^{\dag} 
\end{array}
\right)
= 
\left(
\begin{array}{cccccc}
\Delta & 0 & 0 & 0 & ig_{1} & ig_{1}  \\
0 & -\Delta & 0 & 0 & ig_{1} & ig_{1} \\
0 & 0 & \Delta & 0 & -ig_{2} & -ig_{2} \\
0 & 0 & 0 & -\Delta & -ig_{2} & -ig_{2} \\
-ig_{1} & ig_{1} & ig_{2} & -ig_{2} & -\Omega_{\mathrm{m}} & 0 \\
ig_{1} & -ig_{1} & -ig_{2} & ig_{2} & 0 & \Omega_{\mathrm{m}}
\end{array}
\right)
\left(
\begin{array}{c}
a_{1} \\ a_{1}^{\dag} \\ a_{2} \\ a_{2}^{\dag} \\ b \\ b^{\dag} 
\end{array}
\right)
-i
\left(
\begin{array}{c}
\alpha \\ \alpha^{*} \\ 0 \\ 0 \\ 0 \\ 0
\end{array}
\right),
\end{equation}
which has a pair of trivial eigenvalues $\omega_{0} = \pm \Delta$, and two pairs of hybridized eigenmodes
\numparts
\begin{eqnarray}
\label{omp}
\omega^{(+)} = \pm \sqrt{\frac{\Delta^{2} + \Omega_{\mathrm{m}}^{2}}{2} + \sqrt{ \frac{(\Delta^{2} - \Omega_{\mathrm{m}}^{2})^{2}}{4} - 4\left(g_{1}^{2} + g_{2}^{2}\right)\Delta\Omega_{\mathrm{m}} }}, \\
\label{omm}
\omega^{(-)} = \pm \sqrt{\frac{\Delta^{2} + \Omega_{\mathrm{m}}^{2}}{2} - \sqrt{ \frac{(\Delta^{2} - \Omega_{\mathrm{m}}^{2})^{2}}{4} - 4\left(g_{1}^{2} + g_{2}^{2}\right)\Delta\Omega_{\mathrm{m}} }}.
\end{eqnarray}
\endnumparts
In accordance with \cite{Bernier2018}, for  $\Delta > 0$, $\omega^{(\pm)}$ demonstrate level attraction in the region where $(\Delta^{2} - \Omega_{\mathrm{m}}^{2})^{2} < 16(g_{1}^{2} + g_{2}^{2})\Delta\Omega_{\mathrm{m}}$, which is related to the parametric instability in cavity optomechanics \cite{Aspelmeyer2014}. This region is bounded by the exceptional points, where the eigenmodes coalesce, and the matrix on the right-hand side of (\ref{eom3}) is characterized by the Jordan form:
\begin{equation}
\left(
\begin{array}{cccccc}
-\Delta & 0 & 0 & 0 & 0 & 0 \\
0 & \Delta & 0 & 0 & 0 & 0 \\
0 & 0 & -\sqrt{\frac{\Omega_{\mathrm{m}}^{2} + \Delta^{2}}{2}} & 1 & 0 & 0 \\ 
0 & 0 & 0 & -\sqrt{\frac{\Omega_{\mathrm{m}}^{2} + \Delta^{2}}{2}} & 0 & 0 \\ 
0 & 0 & 0 & 0 & \sqrt{\frac{\Omega_{\mathrm{m}}^{2} + \Delta^{2}}{2}}  & 1  \\
0 & 0 & 0 & 0 & 0 & \sqrt{\frac{\Omega_{\mathrm{m}}^{2} + \Delta^{2}}{2}}
\end{array}
\right).
\end{equation}
For $ \Delta <0 $, in contrast, we have usual mode repulsion.  Remarkably, the energy-level picture for the dispersion relations (\ref{omp}, \ref{omm}) shown in Fig.~\ref{fig2} (b) becomes similar to that in Fig.~\ref{fig1} (b) for the three-oscillator model with inverted horizontal axis.

Solutions of equations of motion (\ref{eom3}) for the cavity modes have the following form
\begin{equation}
\label{res2}
a_{1} = \frac{i\alpha}{\Delta} \frac{\Delta \Omega_{\mathrm{m}} + 4 g_{2}^{2}}{\Delta \Omega_{\mathrm{m}} + 4 (g_{1}^{2} + g_{2}^{2})}, \quad
a_{2}  = \frac{4i \alpha g_{1}g_{2}}{\Delta[\Delta \Omega_{\mathrm{m}} + 4 (g_{1}^{2} + g_{2}^{2})]},
\end{equation}
where we used real $\alpha$ for simplicity.  If $g_{1} \sim \sqrt{n_{2}}$ is nonzero, driving of $a_{1}$ excites also $a_{2}$. In this situation, if we take into account the dissipation of the cavity mode, $\Delta \to \Delta + i\kappa_{\mathrm{c}}$, the response in (\ref{res2}) has $2\pi$-phase shift with respect to the drive due to the degeneracy of the cavity modes with different polarizations.

\section{Aharonov-Casher effect}

Here, we discuss another mechanism of mode attraction for a coupled magnon-photon dynamics inside a ferromagnetic insulator based on the Aharonov-Casher effect \cite{Aharonov1984}.  This effect is related to a dualism between the electrodynamics of electric charges in a magnetic field and that of charge-neutral magnetic dipoles in an electric field. When applied to magnon dynamics this leads, for example, to the Landau quantization of the magnon states under applied electric field gradient \cite{Nakata2017}.  Since the diamagnetism of a conventional electron gas follows from Landau quantization of electron motion, by analogy, we expect that a sort of `diamagnetic' response should also exist for magnons in the electric field. which, In turn, this provides us with a mechanism for the level attraction, as explained below.  

For this purpose, we consider the following spin Hamiltonian for a small ferromagnetic specimen
\begin{equation}\fl
\label{Htot2}
\mathcal{H} = -\sum_{\langle ij \rangle} \left[ \frac{J}{2}\left(S_{i}^{(-)}S_{j}^{(+)}e^{i\theta_{ij}} + S_{i}^{(+)}S_{j}^{(-)}e^{-i\theta_{ij}}\right) + JS_{i}^{(z)}S_{j}^{(z)}\right] - K\sum_{i} \left[S_{i}^{(z)}\right]^{2},
\end{equation}
where $J$ is the ferromagnetic exchange constant, $K$ is the anisotropy parameter, and the summation is over the nearest neigbouring sites limited by the size of the specimen.  The first term contains a field phase associated with the cavity electric field.  The phase factor is given by \cite{Nakata2017}
\begin{equation}
\theta_{ij}(t) = \frac{g\mu_{B}}{\hbar c^{2}} \int_{\bm{r}_{i}}^{\bm{r}_{j}} \left[\bm{E}(\bm{r},t) \times \hat{\bm{z}}\right] \cdot d\bm{\ell},
\end{equation}
where the integration is taken along the path connecting $i$th and $j$th sites, and $z$ plays the role of a spin quantization axis.

\begin{figure}
	\centerline{\includegraphics[width=0.75\textwidth]{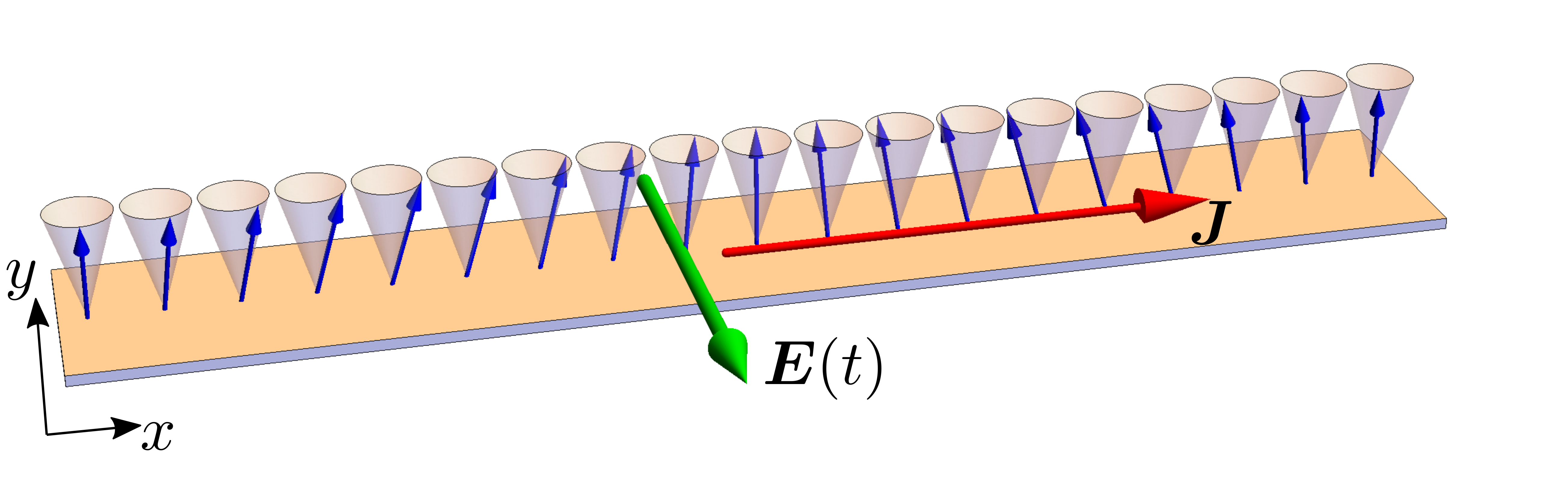}}
	\caption{Schematic picture of a spin wave inside a small ferromagnetic specimen carrying spin current $\bm{J}$ along the $x$ direction interacting with the cavity electric field $\bm{E}$ polarized along the $y$ axis. The electric field is supposed to be homogeneous on the specimen's length scale.}
	\label{fig3}
\end{figure}

If we suppose that the characteristic wavelength of the electric field is much larger than the sample size, then the interaction part in the Hamiltonian (\ref{Htot2}) can be written in the form $-\bm{E}\cdot\bm{J}$, where $\bm{J}=[ig\mu_{B}J/(2\hbar c^{2})]\sum_{\langle ij \rangle}(S_{i}^{(-)}S_{j}^{(+)} - S_{i}^{(+)}S_{j}^{(-)})[\hat{\bm{z}}\times\bm{r}_{ij}]$ is the spin current in the sample (see Fig.~\ref{fig3}).  By quantizing the spin operators, $S_{i}^{(+)} = \sqrt{2S}b_{i}$, $S_{i}^{(-)} = \sqrt{2S}b_{i}^{\dag}$, and choosing the electric field in the form of a linearly polarized cavity standing wave along the $x$ direction, $E_{y}(x) = [\hbar\omega_{c}/(\varepsilon_{0}V)]^{1/2}\sin(\pi n x/L_{x})(a + a^{\dag})$, the interaction energy can be written as $-\hbar\sum_{\bm{k}} g_{\bm{k}}b_{\bm{k}}^{\dag}b_{\bm{k}}(a + a^{\dag})$, where the interaction constant $ g_{\bm{k}} = -[2g\mu_{B}JS/(\hbar^{2} c^{2})][\hbar\omega_{c}/(\varepsilon_{0}V)]^{1/2}\sin(\pi n x_{0}/L_{x})\sum_{\bm{\delta}} \sin(\bm{k}\cdot\bm{\delta})(\hat{\bm{z}}\times\bm{\delta})_{y} $ generally depends on the position of the sample $x_{0}$ with respect to the electric field.

If the sample is populated into a state with $n_{\bm{k}_{0}} \gg n_{-\bm{k}_{0}}$ for some $\bm{k}_{0}$, where $n_{\bm{k}} = \langle b^{\dag}_{\bm{k}}b_{\bm{k}} \rangle$ is the magnon number, and magnon states with different $\bm{k}$ are well separated, we can consider a simplified model
\begin{equation}
\label{Htot3}
\mathcal{H} = \hbar \omega_{c}a^{\dag}a + \hbar\Omega_{\bm{k}_{0}} b_{\bm{k}_{0}}^{\dag}b_{\bm{k}_{0}} - \hbar g_{\bm{k}_{0}} b_{\bm{k}_{0}}^{\dag}b_{\bm{k}_{0}} (a + a^{\dag}) + \mathcal{H}^{(\mathrm{m})}_{\mathrm{dr}},
\end{equation}
where we added the driving term $\mathcal{H}^{(\mathrm{m})}_{\mathrm{dr}} = i\hbar\sqrt{\kappa_{\mathrm{ex}}} \beta_{\bm{k}_{0}}^{(\mathrm{in})}b_{\bm{k}_{0}}\exp(-i\omega t) + \mbox{H.c.}$, with the amplitude of the driving field $\beta_{\bm{k}_{0}}^{(\mathrm{in})}$. The Hamiltonian (\ref{Htot3}) can be considered as dual to the model with magneto-optical interactions in equations (\ref{Hmo}, \ref{Htot1}).  It can be interpreted in terms of the pressure force created by the magnon flow with finite $\bm{k}_{0}$ on the cavity photon oscillator, with the position operator proportional to $a + a^{\dag}$.

Similar to optomagnonic Hamiltonian (\ref{Htot1}), equation (\ref{Htot3}) can be transformed to the time-independent frame by applying the unitary transformation to the magnon operators, $b_{\bm{k}_{0}} \to \exp(b_{\bm{k}_{0}}^{\dag} b_{\bm{k}_{0}}t) b_{\bm{k}_{0}}$ \cite{Aspelmeyer2014}. In this representation, the magnon frequency is replaced by the magnon detuning parameter, $\Omega_{\bm{k}_{0}} \to -\tilde{\Delta}$, where $\tilde{\Delta} = \omega - \Omega_{\bm{k}_{0}}$, with which we can reach the attraction regime in the region $(\tilde{\Delta} - \omega_{c})^{2} < 4 g_{\bm{k}_{0}}^{2} n_{\bm{k}_{0}}$.  We note that, in contrast to optomagnonic case, the single magnon coupling constant is small compared to $\Omega_{\bm{k}_{0}}$.  For realistic parameters, $J = 100$~K,  $\omega_{c} = 10^{12}$~s$^{-1}$, and $S = 1$, we estimate $g_{\bm{k}_{0}} = 10^{-4}$~s$^{-1}$, which means that the effect is difficult to realize experimentally.  However, we believe that it can be relevant to ultrafast magnetization dynamics in metamaterials with $S \gg 1$, and can provide an experimental evidence for the spin current electrodynamics proposed in \cite{Aharonov1984}.

\section{Summary}

We considered several mechanisms of mode attraction for a coupled magnon-photon system inside a microwave cavity.  Using the Maxwell's equations of continuous media, we demonstrated how the level attraction can be described in terms of response functions of the medium.  In particular, we derived the effective permittivity for the inverse Faraday effect, and showed that it can be interpreted as an effective diamagnetic effect in the equations of motion for the electromagnetic fields inside the cavity.  This approach has been supported by the quantum picture based on the optomagnonic Hamiltonian \cite{Kusminskiy2016,Liu2016}, which has reasonable parameter values for experimental demonstration of the effect.

Also, we discussed another mechanism, which is based on the electro-dipole effect of the magnon spin current \cite{Meier2003}.  In the context of level attraction, this mechanism can be considered as dual to optomagnonic approach because it requires driving of the magnon mode with finite wave vector. This effect is relativistically small compared to the magneto-optical mechanism.

\ack
The authors thank Can-Min Hu and Michael Harder for stimulating discussions. 
I.P. is supported by the Ministry of Education and Science of the Russian
Federation, Grant No. MK-1731.2018.2, and by the Russian Foundation for Basic
Research (RFBR), Grant No. 18-32-00769.  R.L.S. acknowledges the support of the
Natural Sciences and Engineering Research Council of Canada (NSERC) RGPIN 05011-18.  The work of R.~Mac\^{e}do is supported by the Leverhulme Trust.

\appendix

\section{Resonant frequencies for the three-oscillator model}
\label{AppA}

In order to derive the Hamiltonian (\ref{Htot}), we consider the magnetization $\bm{M}(t)$ in the static magnetic
field $\bm{H}_{0}$ along the $z$-direction interacting with the microwave field
$\bm{h}(t)$ [see Fig.~\ref{fig1} (a)].  By splitting the magnetization into static and
dynamic parts, $\bm{M} = [M_{s} - m^{2}/(2M_{s})]\hat{\bm{z}} + \bm{m}(t)$,
where $M_{s}$ is the saturation magnetization and $\bm{m}(t)$ is the transverse
dynamical component, the magnetic part of the energy, $\mathcal{H}_{\mathrm{m}}
= -\bm{M} \cdot \bm{H}_{0} - \bm{h} \cdot \bm{M}$, in the linear approximation,
becomes
\begin{equation} \label{Hm}
\mathcal{H}_{\mathrm{m}} = \hbar\Omega_{\mathrm{m}} b^{\dag}b 
- \frac{M_{s}}{\sqrt{2}}\left(h^{(-)}b + h^{(+)}b^{\dag}\right),
\end{equation}
where $\Omega_{\mathrm{m}} = \hbar^{-1}M_{s}H_{0}$, $h^{(\pm)} = h_{x} \pm
ih_{y}$, and the circular components of $\bm{m}$ are expressed via the
Holstein-Primakoff boson operators, $m^{(+)} = \sqrt{2}M_{s}b$ and $m^{(-)} =
\sqrt{2}M_{s}b^{\dag}$. We take the resonant cavity mode in the form of a plane
wave along the $z$-direction, so that the magnetic field is quantized as
follows:
\begin{equation}
\bm{h}(z,t) = i\sum_{\lambda} \left(\frac{\hbar\omega_{c}}
{\mu_{0}V}\right )^{\frac{1}{2}}\cos\left(\frac{\pi N_{c} z}{L_{z}}
\right)\left[a_{\lambda}(\hat{\bm{z}} \times \bm{e}_{\lambda}) 
- a_{\lambda}^{\dag}(\hat{\bm{z}} \times \bm{e}_{\lambda}^{*})\right],
\end{equation}
where $\omega_{c}$ and $N_{c}$ denote the frequency and the
index of the cavity mode, $V$ is the volume of the cavity, $L_{z}$ is the size
of the cavity along the $z$-axis, and $a_{\lambda}$ ($a_{\lambda}^{\dag}$) is
the photon creation (annihilation) operator in the helicity basis
$\bm{e}_{\lambda} = (\lambda, i, 0)/\sqrt{2}$ with $\lambda = 1$ ($-1$) for the
right (left) polarized wave.  Using this quantization of the magnetic field in
(\ref{Hm}), and adding the electromagnetic energy of cavity photons, we
obtain the Hamiltonian (\ref{Htot}) with $g_{0} = (\omega_{c}M_{s}^{2}/\hbar
\mu_{0}V)^{1/2}$ if the sample is at the position of maximum magnetic field.

The Heisenberg equations of motion for the Hamiltonian (\ref{Htot}) with the driving have the following form:
\begin{eqnarray}
-i\dot{a}_{R} &= -\omega_{c} a_{R} - g_{0}b^{\dag} - i\sqrt{\kappa_{\mathrm{ex}}} \alpha_{\mathrm{in}} e^{-i\omega t}, \\
-i\dot{b}^{\dag} &= \Omega_{\mathrm{m}} b^{\dag} + g_{0} a_{R} + g_{0} a_{L}^{\dag}, \\
-i\dot{a}_{L}^{\dag} &= \omega_{c} a_{L}^{\dag} + g_{0} b^{\dag}.
\end{eqnarray}
The stationary response of the system to the driving at frequency $\omega$ can be found from the algebraic system of equations for the field amplitudes 
\begin{equation} \label{eom}
\left(
\begin{array}{ccc}
\omega - \omega_{c} & -g_{0} & 0 \\
g_{0} & \omega + \Omega_{\mathrm{m}} & g_{0} \\
0 & g_{0} & \omega + \omega_{c}
\end{array}
\right)
\left(
\begin{array}{c}
a_{R} \\ b^{\dag} \\ a_{L}^{\dag}
\end{array}
\right)
= i\sqrt{\kappa_{\mathrm{ex}}}
\left(
\begin{array}{c}
\alpha_{\mathrm{in}} \\ 0 \\ 0
\end{array}
\right),
\end{equation}
which has the following solutions
\begin{eqnarray}\label{res}
a_{R} &= \frac{i\sqrt{\kappa_{\mathrm{ex}}}\alpha_{\mathrm{in}}\left[(\omega + \omega_{c})(\omega + \Omega_{\mathrm{m}}) - g_{0}^{2}\right]}{\left(\omega^{2} - \omega_{c}^{2}\right)(\omega + \Omega_{\mathrm{m}}) + 2 g_{0}^{2} \omega_{c}}, \\
b^{\dag} &=  \frac{i\sqrt{\kappa_{\mathrm{ex}}}\alpha_{\mathrm{in}} g_{0}(\omega + \omega_{c})}{\left(\omega^{2} - \omega_{c}^{2}\right)(\omega + \Omega_{\mathrm{m}}) + 2 g_{0}^{2} \omega_{c}}, \\ \label{res1}
a_{L}^{\dag} &= \frac{i\sqrt{\kappa_{\mathrm{ex}}}\alpha_{\mathrm{in}} g_{0}^{2}}{\left(\omega^{2} - \omega_{c}^{2}\right)(\omega + \Omega_{\mathrm{m}}) + 2 g_{0}^{2} \omega_{c}}.
\end{eqnarray}

The resonant frequencies can be obtained from the determinant of the matrix on the left hand side of equation (\ref{eom}), $(\omega^{2} - \omega_{c}^{2})(\omega + \Omega_{\mathrm{m}}) + 2 g_{0}^{2} \omega_{c} =0$.
The solutions have the form
\begin{eqnarray}
\omega_{0} & = -\frac{1}{3}\left[\Omega_{\mathrm{m}} + \left(A - iC\right)^{\frac{1}{3}} + \left(A + iC\right)^{\frac{1}{3}}\right], \\
\omega_{1} & = -\frac{1}{3}\left[\Omega_{\mathrm{m}} - e^{i\pi/3}\left(A - iC\right)^{\frac{1}{3}} - e^{-i\pi/3}\left(A + iC\right)^{\frac{1}{3}}\right], \\
\omega_{2} & = -\frac{1}{3}\left[\Omega_{\mathrm{m}} - e^{-i\pi/3}\left(A - iC\right)^{\frac{1}{3}} - e^{i\pi/3}\left(A + iC\right)^{\frac{1}{3}}\right],
\end{eqnarray}
where $A = \Omega_{\mathrm{m}}^{3} -9\Omega_{\mathrm{m}}\omega_{c}^{2} + 27g_{0}^{2}\omega_{c}$ and $B=\Omega_{\mathrm{m}}^{2} + 3\omega_{c}^{2}$, and $C = \sqrt{B^{3} - A^{2}}$.  The frequencies are real in the region where $B^{3} \ge A^{2}$.  Otherwise, $\omega_{1}$ and $\omega_{2}$ become complex.  In the limit $g_{0} = 0$, we have $\omega_{0} = -\Omega_{\mathrm{m}}$, $\omega_{1} = -\omega_{c}$, and $\omega_{2} = \omega_{c}$.

\section{The effective permittivity for the inverse Faraday effect}
\label{AppB}

In order to derive the effective response for the inverse Faraday effect, we consider the electric displacement field
\begin{equation}
\delta D_{i}(t) = i\varepsilon_{0} f \epsilon_{ijx} m_{x}(t) E_{j}(t).
\end{equation}
In the Fourier space, this expression can be written as
\begin{equation} \label{Dw}
D_{i}(\omega) = i\varepsilon_{0}f\epsilon_{ijx}\int_{-\infty}^{\infty} \frac{d\omega'}{2\pi} \chi_{xx}(\omega') H_{x}^{\mathrm{eff}}(\omega') E_{j}(\omega - \omega'),
\end{equation}
where the magnetization dynamics in the effective field is taken into account through the ferromagnetic susceptibility,  $m_{x}(\omega) = \chi_{xx}(\omega)H_{x}^{\mathrm{eff}}(\omega)$, and the Fourier components of the fields are determined as follows $F(\omega) = \int_{-\infty}^{\infty} dt e^{i\omega t} F(t)$. The Fourier component of the effective field is given by 
\begin{equation}
\bm{H}^{\mathrm{eff}} = \frac{i\varepsilon_{0}f}{4}\int_{-\infty}^{\infty} \frac{d\omega'}{2\pi} \bm{\mathcal{E}}(\omega') \times \bm{\mathcal{E}}^{*}(\omega - \omega'),
\end{equation}
which gives after substitution into (\ref{Dw})
\begin{eqnarray} \label{Dw1}
D_{i}(\omega) = -\frac{\varepsilon_{0}^{2}f^{2}}{8} \epsilon_{ijx} \epsilon_{klx}\int_{-\infty}^{\infty} \frac{d\omega'}{2\pi} \int_{-\infty}^{\infty} \frac{d\omega''}{2\pi} \chi_{xx}(\omega') \mathcal{E}_{k}(\omega'')\mathcal{E}_{l}(\omega' - \omega'') 
\nonumber
\\ 
\times \left[\mathcal{E}_{j}(\omega - \omega') + \mathcal{E}_{j}^{*}(\omega - \omega')\right].
\end{eqnarray}

We imply that the electric field is driven by the external field $\bm{E}_{0}(t) = [\bm{\mathcal{E}}_{0}\exp(-i\omega_{\mathrm{L}}t) +  \bm{\mathcal{E}}_{0}^{*}\exp(i\omega_{\mathrm{L}}t)]/2$, and expand (\ref{Dw1}) up to the linear order in the fluctuating field, $\bm{E} = \bm{E}_{0} + \delta \bm{E}$, which gives after some algebra
\begin{eqnarray}
\delta D_{i}(\omega) = -\frac{\varepsilon_{0}^{2}f^{2}}{8} \epsilon_{ijx} \left\lbrace 
\chi_{xx}(0) \left(\bm{\mathcal{E}}_{0} \times \bm{\mathcal{E}}^{*}_{0}\right)_{x} \left[\delta\mathcal{E}_{j}(\omega) + \delta\mathcal{E}_{j}^{*}(\omega)\right] 
\right. 
\nonumber
\\
\left.
+ \chi_{xx}(\omega - \omega_{\mathrm{L}}) \mathcal{E}_{0j}\left[\mathcal{E}_{0z}^{*}\delta\mathcal{E}_{y}(\omega) - \mathcal{E}_{0y}^{*}\delta\mathcal{E}_{z}(\omega)\right] 
\right. 
\nonumber
\\
\left.
+ \chi_{xx}(\omega + \omega_{\mathrm{L}}) \mathcal{E}_{0j}^{*}\left[\mathcal{E}_{0y}\delta\mathcal{E}_{z}^{*}(\omega) - \mathcal{E}_{0z}\delta\mathcal{E}_{y}^{*}(\omega)\right]
\right. 
\nonumber
\\
\left.
+ \chi_{xx}(\omega - \omega_{\mathrm{L}}) \mathcal{E}_{0j}\left[\mathcal{E}_{0y}\delta\mathcal{E}_{z}^{*}(\omega - 2\omega_{\mathrm{L}}) - \mathcal{E}_{0z}\delta\mathcal{E}_{y}^{*}(\omega - 2\omega_{\mathrm{L}})\right] 
\right.
\nonumber
\\
\left.
+ \chi_{xx}(\omega + \omega_{\mathrm{L}}) \mathcal{E}_{0j}^{*}\left[\mathcal{E}_{0z}^{*}\delta\mathcal{E}_{y}(\omega + 2\omega_{\mathrm{L}}) - \mathcal{E}_{0y}^{*}\delta\mathcal{E}_{z}(\omega + 2\omega_{\mathrm{L}})\right]
\right\rbrace.
\end{eqnarray}
Neglecting the last terms with $\omega \pm 2 \omega_{\mathrm{L}}$ and the first term, which corresponds to the static component of the effective field, we obtain equation (\ref{dia}).

\vspace*{1em}
\bibliographystyle{unsrt}
\bibliography{attraction}

\end{document}